# Design and implementation of a Framework for remote experiments in education

Pavel Kuriščák
Charles University, Faculty of Mathematics and Physics,
Laboratory of General Physics Education
Prague, Czech Republic
kurispav@fjfi.cvut.cz

Pedro Rossa
Universidade de Lisboa,
Instituto Superior Técnico,
Instituto de Plasmas e Fusão Nuclear
Lisbon, Portugal
pedro.rossa@tecnico.ulisboa.pt

Horácio Fernandes
Universidade de Lisboa,
Instituto Superior Técnico,
Instituto de Plasmas e Fusão Nuclear
Lisbon, Portugal
hf@ipfn.ist.utl.pt

João Nuno Silva
Universidade de Lisboa,
Instituto Superior Técnico,
INESC-ID Lisboa
Lisbon, Portugal
joao.n.silva@inesc-id.pt

*Abstract*— Remote Controlled laboratories is a teaching and learning tool that increasingly becomes fundamental in the teaching and learning processes at all the levels. A study of available systems highlights a series of limitations on the used programming languages, overall architecture and network communication patterns that, that hinder these systems to be further adopted. Current technologies and modern WEB architectures allow the resolution of such limitations.

Here we present the FREE (Framework for Remote Experiments in Education) platform, a novel system, that, using modern technologies, architectures, and programming practices, will be easier to integrate with external tool and services and new experiments.

FREE was developed in Python, Django programming framework, HTML, JavaScript, and web services to easy the development of new functionalities. The designed architecture provides a louse coupling between the infrastructure and the remote experiments facilitating further developments and allow new experiment integrations.

Currently FREE is already running in various countries providing access to about five types of experiments in the area of physics), integration with various Learning Management Systems and external Authentication mechanisms. Using FREE the development and integration of new experiments (independently of the supporting Hardware and programming language) is now easier to be made available to remote users.

## I. Introduction

Remote controlled laboratories (RCL) have been used in teaching various fields of engineering and sciences but COVID-19 and the requirement for remote learning has given a boost to their use and a visibility for its need and usefulness.

With RCLs students can access laboratory apparatus and execute experiments deploying and configured in remote locations. The access to these remote controlled laboratories is performed using regular PC either using specific desktop applications or a simple web browser. The experimental apparatus can be configured before the start of the experiment execution or while the experiment is executed. On the physical location where the experiment is installed, a server allows and controls access to the physical apparatus.

The access to remote experiments further enriches the students' learning, by providing access to resources and practical knowledge that otherwise was inaccessible. The use of a controlled interface for the setup of the experiments also facilitates the operation and the knowledge of the various experiment parameters.

Remote controlled laboratories or single experiments were already used before COVID but the pandemic and the resulting teaching and learning adjustment made this type of learning activity more relevant. Even with the return to normal classes the need and advantages of the use of remote controlled laboratories remains.

Most of the current RCLs provide access to experimental apparatus that are not available or easily operated by schools due to cost and operation specifications, but some of the widely available experiments that students access in schools' laboratories can be easily converted to allow the remote access. This adaptation not only facilitates the executions (in space and time) but also allows its integration into Learning Management Systems such as Moodle or Canvas.

This possible widespread use of remote experiments is in part limited by the available computational infrastructures to deploy and make RCLs available. A simple survey on the existing remote controlled laboratories infrastructures proves these systems' architecture and implementation technologies make the widespread use and addition of experiments a challenging task. Some of the systems are monolithic with the available experiments hardcoded into the system, some others require the installation of specific software on the students' computers hindering its use.

With these challenges in hand, we present here a new open architecture for the implementation of remote controlled laboratories or experiments: the Framework for Remote Experiments in Education (FREE) architecture. Since FREE is being developed from the ground up, some requirements, not necessarily functional, drive the project decisions. One of the main requirements is the ease of integration of new experiments. This should be accomplished by the definition of a clear and easily replicable code structure, a clear separation between components and a set of well-defined Interfaces (network and programmable). The code structure will allow a simple development of the FREE components that will implement the User Interface, the clear separation of components and the interface will allow the use of the most suitable programming languages for the implementation of the experiments. The second requirement is the possibility to integrate the developed systems (and available experiments) into external services, such as remote authentication, use in LMS, or even federated authentication and authorization. This requires the adoption of a programming framework that provides extensible and suitable libraries.

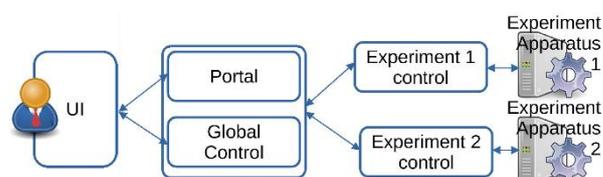

*Figure 1 - RLC generic architecture*

A FREE system is a web application running on a server, the user interface is accessed through a web browser and the experimental apparatus are remote processes. The openness in the proposed system comes from the definition of a set of software components that allows the simple definition of new experiments and a set of well-defined interfaces (both for programming the interactions and perform communication). The while communication between the UI (User Interface), Server and experimental apparatus is based on HTTP, REST and Json.

The FREE architecture is already implemented in a system that allows access to remote experiments. It is extensible with respect to the number and type of experiments and integration with various authentication mechanisms. Users can access various remote experiments using a regular browser and accessing a single server. After authentication with one of the widely available services (for instance Google or Microsoft), the user can configure and execute any of the available experiments. During the experiment execution, a video feed and the results are shown to the student in real time. After the execution the user can access and download the relevant results.

The implementation and addition of new experiments is facilitated by the modularity of the User Interface and the communication protocol between the FREE and the experimental apparatus. The definition of the interface is accomplished by the implementation of simple HTML whose input fields correspond to the parameters of each execution. The steps to verify the values and control the executions are generic to all the types of experiments. The FREE interacts with the experimental apparatus using simple communication protocol. The apparatus polls the FREE for new execution, and, when a new execution is to be started, the data retrieved from the HTML interface is transmitted to the apparatus. During the experiment execution, the apparatus sends partial results to the FREE server that forwards them to the web page.

The current version of FREE was implemented in python and several types of experiments are now ported to it. The programmers of the new experiments write minimal HTML and JavaScript (for the interaction with the user) and only need to implement simple Request-Reply protocols (with a limited number of interactions) on the experiment apparatus.

The remaining of the paper is organized as follows, Section II presents and describes existing systems, highlighting their main functionalities, and limitations. The following chapter describes the requirement architecture and implementation of the proposed RCL infrastructure. Chapter IV presents and validates the usage of FREE. The document terminates with a conclusion and brief description of future work and development possibilities only possible due to FREE.

## II. REMOTE CONTROLLED LABORATORIES

The main innovation of the work presented here relates to the overall software infrastructure and how it affects the integration of these experiments into Learning Management Systems (LMS) or external authentication services, the cost to support and develop new features or even how to add new types of experiments.

### A. Publicly available remote controlled laboratories

The major work in RCLs has been done in the engineering areas, but there are some systems applied to life sciences. Most of the publicly available laboratories provide experiments in the areas of physics, mechanical, electrical engineering, or even computer science [1].

These laboratories and experiments use various actuators to control physical processes and sensors to retrieve relevant results. The generic architecture includes a user interface implemented using web technologies, so that a user with a web browser and internet access can configure/control the experiment and retrieve the results.

Currently some remote experiments are available for any user to access thorough the internet. Any user can configure, run them, and see the results in the following platforms: LabsLand [2], RELLE [3], ISE [4], elab [5] or Unilabs [6]. Most of these remote controlled laboratories list the available experiments on a simple web page. Only a few aggregates the experiments into a portal (LabsLand, elab, and Unilabs).

In LabsLand the authentication can be provided by a school after which the user has full access to the selected experiments, or using a trial google account supplying access to a limited number of experiments executions. In elab, users can use the authentication provided by two Moodle installations or run the experiments as an anonymous user. In Unilabs users can create a free account to access the Moodle contents and the available experiments. In all other, users have full access to all the available experiments.

In most systems, each remote experiment is presented in a web page, where the user controls a physical apparatus to execute the experiment. The way the user controls the physical experiments differs a bit between the various systems, but follow the generic experiment control [7] pattern: define experiment parameters, start experiment, see experiment video stream, see results.

In some systems the user must previously reserve a time slot in which the user has exclusive control of the experiment. Most systems use a web-browser to allow users to control the experiment. Elab is the only system that requires the user to download a Java application that will present the user interface. Besides the control and visualization of the experiment, only in elab are the users allowed to download the results in a tabular (csv file) form. In all other systems the user can only see a plot with the results.

Unilabs offers a set of remote and virtual laboratories, mediated by Moodle. The access to the experiments is done navigating a Moodle site. Some experiments are freely accessible without registration, while others are integrated in a course, requiring the user to create an account and enroll in a course.

*B. Computational infrastructures*

The work here presented is focused more on the supporting infrastructure than on the functionalities offered to the users, so it is important to understand how the architectural and technological decisions made on exiting systems affect the quality of the supporting infrastructure. This section will present a detailed view of the internals of the elab, RELLE [8], WebLab-Deusto [9] (supporting infrastructure of LabsLand), REMLABNET [10] supporting library (used to implement ISE) and other isolated research works [11, 12, 13, 14]. Unilabs is based on Moodle modules, and although there is no specific documentation to implement a new remote experiment, it is possible to infer its architecture and programming requirements.

*1) Architecture*

The various supporting software systems presented here follow a generic architecture with 4 main components: User Interface (**UI**), **Portal**, the **Global Control**, and the specific **Experiment Control**, as illustrated in Figure 1

The **Portal** and **Global Control** usually runs on a server and manages the various available experiments, the user authentication and authorizations, and execution queues and reservations. **Experiment Control** is the low-level code that controls the physical experiment, sets it up, starts execution, retrieves the sensor data, and sends it to the server. The **UI** presents the user with the available experiments (provided by the **Portal**) and shows the **Experiment Control** and execution results.

The various supporting systems implement this architecture in diverse ways, and with different usability levels. The RELL implements one variation of this generic architecture, with no **Global Control**. Right after the user selects one experiment on the **Portal**, all the interaction is done directly to the **Experiment Control**. This part is running in the microcontroller/raspberry pi directly attached to the physical experiment. The REMLABNET does not implement the **Portal** and **Global Control** has limited functionalities. Each experiment is a different implementation of the **UI**, **Global Control** and **Experiment Control**. The **UI** is implemented using WEB technologies and the experiments control runs on a different computer from **Global Control**. REMLABNET allows two different interactions between user **UI** and the experiments.

In REMLABNET and RELLE, regular users' browsers connect directly to the computer attached to the physical experiment. This solution is simpler than one with a **Global Control** module, but it is less secure and reduces the integration and services offered to the users. Only in isolated experiments [12, 13] this type of architecture can be practical.

Both WebLab-Deusto and elab implement the 4 components and distribute them in the best way. The **UI** executes in the user computer, both the **Portal** and **Global Control** execute on the server and each Experiment has its own **Experiment Control**. The **Experiment Control** can run on a different computer attached to the Experiments and communicated with the server through a specific API/message. elab implements the **Portal** using Web technologies, allowing its access using a regular browser, but to execute any experiment it is necessary to install a Java application. In the WebLab-Deusto **Experiment Control** 1 can be hosted inside the whole infrastructure not requiring a different computer or processes.

Unilabs is supported by Moodle that offers some of the **Portal** and **Global Control** functionalities. The **UI** of Unilabs is web based. Since there is no specification for the **Experiment Control** modules, these can be placed in remote computers of local to the Moodle server, depending on the particularities and experiment deployment.

*2) Communication protocols*

To allow remote access to the physical experiments these components are distributed in diverse ways and communicate between them using different network protocols.

In RELLE and REMLABNET the communication between the user computer and the **Experiment Control** simply uses HTTP and HTML. The Experient Control, the computer or raspberry Pi directly attached to the physical experiment, should implement a regular web server to supply the **UI** and answer the user requests.

Elab uses CORBA [15] to implement communication between the various components: **UI**, **Global Control** and **Experiment Control**. WebLab-Deusto uses regular HTTP/HTML to perform communication between the browser and the server. The interaction between the server and the **Experiment Control** is done using HTTP/REST methods. In these two systems the **Experiment Control** (running close to the experiment) receives requests and commands from the server. Since it is the server that contacts the **Experiment Control** on a remote computer, the server needs to know the exact network location of the **Experiment Control**. This solution, as in the RELLE and REMLABNET, may require complex network configuration.

For Unilabs there is no specific protocol to be used, so the communication between the Server (g **Global Control**) and the **Experiment Control** (if existing in a remote computer) can be implemented using any Internet base technology or protocol and be different for each experiment.

*3) Programming language and environment*

elab uses JAVA as the main language for the **Global Control**, **UI**, and **Experiment Control**, and uses the language supporting features for the graphical elements. The communication between components can be programmed using CORBA allowing some sort of programming uniformity between components implemented in different languages. To implement a new experiment, **Experiment Control** can use any language if it supports CORBA

REMLABNET uses either Java or PHP for the development of the Experiments control modules. Since there is no public documentation, it is hard to understand the complexity and effort to integrate or develop new components.

RELLE remote experiments can be implemented using any server-side language: the whole interaction with the user is directly done using the browser and no communication needs to be done with the **Portal**. Experiment registration is done via a web page on the **Portal** part, with the insertion of the experiment data (name, description, type, location, …).

The WebLab-Deusto is implemented in Python using the Flask [16] framework, and the SQL Alchemy [17] for database access. There are two distinct types of experiment implementations: managed and unmanaged. In the case of managed Experiments, the **Experiment Control** is integrated into the server and can be implemented in python or any other programming language that supports XML-RPC. If implemented in python the local method calls are used to interact with the experiment, while to other languages, XML-RPC is used. Unmanaged are external processes that implemented a web service to receive commands from the server. These commands will be generated in the **UI** (JavaScript running in the browser) and be sent to **Experiment Control** or generated by **Global Control**.

Although these commands can be specific for each experiment, there is a set of pre-defined ones:

- do_start_experiment – start running the experiment
- do_dispose – stop experiment and/or clear resources
- do_send_file_to_device – used to configure experiment
- do_send_command_to_device – used to configure experiment
- do_should_finish – retrieves experiment execution status

To implement a new experiment, it is necessary to implement the **Experiment Control** and create specific configuration files before its registration on the server.

The module components for UNILAB should follow the Moodle specifications and be implemented in PHP. The user interface for the experiments control can be implemented in JavaScript [18]. All experiment data are stored in Moodle database, being available for latter processing or use.

Of these five systems, only WebLab-Deusto is open source. Others' documentation doesn't give any information on how to access, use or install their source code.

*4) External services*

Some of these systems are supported by external authentication and Learning Management Systems. These two integrations allow ease the user management and facilitates the users access to the experiments. The Integration with LMS further allows the development of customized exercises dependent on the execution of the experiments.

With respect to authentication, only WebLab-Deusto allows the use of remote authentication services (such as Google or Microsoft). In this system users do not need to create a new account; they just need to use their cloud service credentials. This is an advantage for the users that do not need to define and memorize a new username/password, but also for the remote controlled laboratory administrators that do not need to manage and secure users passwords.

WebLab-Deusto, REMLABNET, and Unilabs are integrated into Moodle, allowing the development of courses that require the execution of remote experiments. WebLab-Deusto and REMLABNET [19] uses LTI [20] to perform the integration with Moodle. Here these systems implement a set of endpoints that are called from the LMS. Users seamlessly access content local to the LMS and the remote experiments.

The other systems require the development of specific MOODLE activity modules. The programmer needs to implement a library and set of classes (following a MOODLE specification) using PHP and JavaScript. In small scale systems such as Laborem [14] this solution may be adequate, but for large scale systems is limiting

*C. Limitations*

From our point of view none of the available systems provides all the necessary requirements and functionalities on a modern, easy to integrate and extend RCL. All of them present some characteristics that can be further improved. Since the first goal was to re-use or implement a new remote controlled laboratory system, these limitations and deficiencies should be found.

The existence of a **Portal** and **Global Control** that mediates all the interaction between the user and the remote experiment is a security requirement that must be implemented. RAMLABNET and RELLE require the client browsers to contact directly with **Experiment Control**. This poses a big security thread for every experiment and requires complex network configuration and verification of firewalls.

With respect to programming languages, it is obvious that nowadays the use of Java (in elab) is a major limitation. With respect to the overall architecture, it is our understanding that **Global Control** and **Experiment Control** should be clearly separated with the possibility to be executed on a different computer. WebLab-Deusto and elab clearly enforce this.

With respect to communication protocols the major drawback of the presented solutions refers to the direction of communication between the remote **Global control** (running on the server) and the **Experiment control**. In WebLab-Deusto and elab it is the server that initiates communication with the **Experiment Control**. This requires that an experiment installed in a different network or institution has a public address (so that the **Experiment Control** can contact it) or a VPN needs to be used. This approach not only requires complex network configuration but also decreases the security of the experiment.

The integration with remote authentication mechanism should also be taken into consideration when supplying a **Portal** to access the experiments. WebLab-Deusto is the only one that offers such functionality. The integration with Moodle using specific modules (such as in Unilabs and Laborem) limits the interoperability with other LMS such as canvas, Sakai or Blackboard.

After a review of the presented system the one that clearly implements the best solutions is WebLab-Deusto, but further detailed evaluation can clearly find its drawbacks and limitations. The direction of communication between the server and the **Experiment Control** and with the programming framework requires complex network configuration if the remote experiment is installed in a different network or institution. It either has a public address or, if behind a NAT, is connected to the server network using a VPN. Although this system uses python (a modern and perfectly suitable programming language) the web framework can be further improved if it implements a more standard programming environment and modality. Django for instance, supplies more consistent project organization and defines a straightforward way to implement reusable modules than Flask. The third limitation of the use of WebLab-Deusto is related to the way the integration is done with LMSs. It uses an added proxy (gateway4labs [21]) that supplies interoperability functionalities by implementing LTI [20], at the expense of an added layer of complexity.

## III. THE FRAMEWORK FOR REMOTE EXPERIMENTS IN EDUCATION

This chapter describes the implementation of the FREE (Framework for Remote Experiments in Education), a novel remote controlled laboratories management systems, by first listing the global requirements that guided the technological decisions. After the presentation of the Architecture and Data Model of our solution, the technical implementation details and decisions are presented.

### A. Requirements

The general functional requirements that FREE should implemented are those common to most of the Remote Controlled Laboratories management systems:

- (R1) it should allow users to control an experiment that is in a remote location,
- (R2) users should use a browser without the need to install any other software,
- (R3) users should have access to various experiments on the same portal,
- (R4) The user should have access to a page where the selected experiment is configured,
- (R5) user should see video and results of the experiment execution in real-time, and
- (R6) users should have access to past results.

With respect to the operation of the experiments system, the following requirements were set:

- (R7) Experiment control (from Figure 1) should not have public address nor be accessible from the Internet.
- (R8) One Apparatus can supply several types of related experiments for the user to execute
- (R9) Replicas of the same experiment can be provided by multiple apparatus installed in multiple locations

Another set of concerns and requirements is related to the integration with external components and programming of new features or experiments

- (R10) FREEE should allow the integration of multiple authentication methods
- (R11) FREE should be interoperable with available LMS
- (R12) Addition of new experiments should not require the programming and changes on the FREE server
- (R13) The programming used in the FREE server should be well defined and standardized
- (R14) The interface between provided by FREE server should be well defined and allow the interoperability with Experience Controls implemented in different programming languages

### B. Architecture and data model

FREE follows the generic architecture presented in Figure 1, with its 4 main components presented in Figure 2: UI is presented to the user through the browser, The server will implement a portal with a list and provide access to all the experiments. The server will also handle the global control of all the experiments (scheduling of execution, access control, results management, …) and the Experiment Control will be placed on a computer close to the Experiment apparatus.

The first significant difference from the existing solutions is the capability of an Experiment Apparatus to be able to execute different Experiments (as illustrated with the Apparatus A that can execute Experiments 1 and 2).

The second fundamental difference (fundamental to fulfill R7) lies in the communication between Experiment Control and the FREE Server. In FREE the server never starts communication with the remote Experiment Control, it is always the other way. This reduces the complexity of network configurations and increases the security of the experiments. This also simplifies the development of Experiment Control (it is a simple client) and its deployment without the need to previously defined the Experiment control network address.

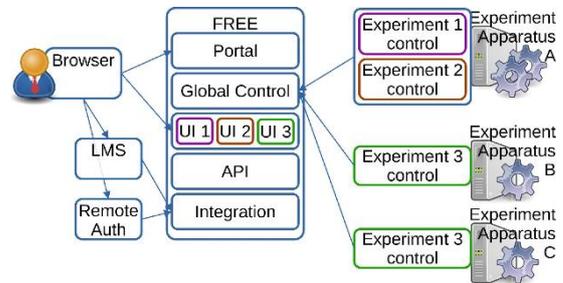

*Figure 2 - FREE detailed architecture*

All the user interaction is programmed using web technologies: a backend server and HTML/JavaScript/REST on the browser. The generic parts of the UI (portal, authentication, list of executions and results) are common to all the experiments, but it is necessary to define specific forms (for the creation of experiments) and results visualization for each experiment. This is carried out with the development and installation of specific modules for each experiment. These modules have the specific HTML and JavaScript that is presented to the user.

In the interaction between the User (UI) and the Experiment Control/Apparatus) the FREE Server acts as a simple Proxy. It receives from the browser the data corresponding to the configuration of the Experiments and forwards it unprocessed to the Apparatus when asked. The flow of the results is similar, during the experiment Execution, the Apparatus sends periodically the results to the FREE Server that stores them unmodified in the database. The JavaScript in browser requests those results, formats and present them to the user. The FREE Server is completely agnostic to the format of the Experiment configuration and Execution results. These can be Json or XML that is produced and processed only by the JavaScript of the UI and by the Experiment Control.

FREE implements two more modules to allow its interaction and integration with the remote components and systems. The API (Application Program Interface) module will allow the Experiment Control processes to retrieve the experiments to be executed and supply status and results. The Integration Module will allow the use of external authentication methods (from Google, Microsoft, or others) and the integration of the existing experiences into LMSs such as Moodle.

Figure 3 presents the data model implemented by the FREE. This diagram represents the database tables of all the necessary information: the various connected apparatus, the available experiments, the executions and results. For simplicity, the specific attributes of each class are not presented here. The data here represented is persistently stored in the database.

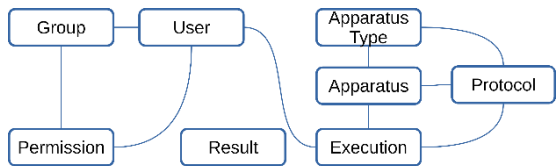

*Figure 3 - FREE Database model*

The top level data types implemented by FREE are the Apparatus Types and Protocols. The apparatus type corresponds to the several classes of physical hardware that can be connected to FREE. For instance, in Figure 2 two distinct types of Apparatus exist: the one that implements the red and maroon experiments and the one that implement the green experiments. The second level datatype is the protocol that allows the representation of the available experiment types. In the previous examples there are 3 different protocols, corresponding to three distinct types of experiments.

This data modeling decision complemented by the decoupling of the Experiment Control from the FREE Server allows each Apparatus (the physical hardware in the laboratory) to implements one or various Protocols and the existence of multiple similar apparatus that implement the same Protocol (requirements R8 and R9). In Figure 2 shows 3 apparatus that implement 3 distinct types of experiments (protocols).

Users can belong to groups that give them access to various permissions of administration or experiment's accesses. When defining an experiment execution, the user selects one Apparatus, one of the Protocols implemented by such apparatus, and defines the initial execution parameters. After the execution is created, the Apparatus will retrieve its configuration, execute it and send to FREE the Results.

Each protocol record will store not only the specific Apparatus configuration but also the initial configuration that should be provided by the user. The Execution and Results records will store all the relevant information about each experiment execution: its type (Protocol) on what Apparatus was executed at, the initial parameters and all the results. Besides this data, FREE also associates to each Protocol and Apparatus the necessary HTML and JavaScript code to implement the UI on the user browsers.

### C. Implementation

To satisfy the requirements R10, R11, R12 and R13, a suitable programming language and framework should be selected. From a study of the existing programming languages and corresponding frameworks the selected ones were Python and Django. The use of this combination provides a set of advantages on the development, maintenance, and evolution of the FREE project.

Python is a widely used language with a large programming base. The number of proficient programmers in this language is continuously growing, guaranteeing the manpower t evolve the system in the future. The selection of Django instead of other options (such as flask) was decided taking into consideration its well defined project structure and the considerable number of available libraries.

The use of a Django also allows the choice of any database backed that best suit the infostructures where FREE will be installed. Different FREE installations will be able to use different database systems (MySQL, Microsoft SQL Server, or any other supported by Django) with minimal FREE configuration. Django also offers simple and efficient mechanisms to upgrade and install new code in existing installations.

*1) API*

To fulfill requirements R7 and R14, it was decided to implement the FREE interface based on REST webservices. These REST webservices are separated into two sets: one for the UI and another for the Apparatus. In FREE, it is the Apparatus that initiates all interaction with the FREE server. Each Apparatus calls the provided endpoints to retrieve experiments and supply results and status. For the UI, besides static pages that show the Apparatus, and protocols static information and forms, the creation of experiments and retrieving on results is done by JavaScript REST calls to the implemented webservice.

The REST API endpoints for the interaction between the browser and FREE server are as follows. For simplicity, the corresponding method is also presented:

- **POST /execution** - Configures an execution of a certain experiment,
- **GET PUT DELETE /execution/<int:id> -** Reads, updates or deletes a given execution,
- **GET /execution/<int:id>/result** - Returns a list of all results for the execution
- **GET /execution/<int:id>/result/<int:last_id>** - Returns a list of the more recent results for a given execution,
- **PUT /execution/<int:id>/start** - Starts a previously configured execution.
- **GET /execution/<int:id>/status** - Retrieves the status of a given execution.

For the interaction between Apparatus and the FREE server, the following REST endpoints are available:

- **PUT /execution/<int:id>/status** - Changes a given execution status
- **PUT /apparatus/<int:id>/heartbeat** - Notifies the system that the Apparatus is alive
- **GET /apparatus/<int:id>/nextexecution** - Returns the next execution that the apparatus should perform.
- **POST /result** - Adds a partial or final result to a given execution,
- **GET /apparatus/<int:id>/queue** - Retrieves the complete list of waiting executions for the apparatus.
- **GET /apparatus/<int:id>** - Retrieves information or configuration about an apparatus.

*Figure 4 - Administrative forms - configuration of a protocol*

This is the minimum set of endpoints that allows the select type of interaction: the users can configure, execute, and see the results of experiments, and the Experiment Control, can retrieve specific configurations, and return to the server the status and results. The code of each of the described endpoints processes, stores, and reads information on the database. Furthermore, to identify remote users and Apparatus and to guarantee Access control modern techniques, such as security token and secrets, are implemented.

In the definition of an experiment configuration and processing of results, the FREE Server is completely transparent to each specific experiment data, it only stores and forwards this information. All the processing of this information is only done in the JavaScript code in the UI and on the Experiment Control next to Apparatus. The use of REST endpoints further simplifies the development and integration of new experiments with no changes to the infrastructure and using different programming languages on the browser or on the Experiment Control (R14). Additional endpoint can be added to implement new functionalities, such as experiments reservations or control during the experiment execution.

*2) Experiments*

To create a new experiment and integrate it into FREE it is necessary to follow three steps: (i) implement the Experiment Control that will call the previously presented endpoints to interact with the FREE server, (ii) define the UI and (iii) insert the relevant data into the FREE server database.

In the creation the UI, the programmer does not need to write any python server code, he only needs to define the HTML elements that will allow the insertion of the experiment configuration and write the JavaScript code to show the results on the web page. This is done in a template file that will hold all the supporting HTML and JavaScript code. The programmer needs to design a simple HTML form (specific for that experiment configuration), and a DIV with the tables and plots to be filled with the results. The

```
FREE_GOOGLE_OAUTH=on
SOCIAL_AUTH_GOOGLE_OAUTH2_KEY= 22218771
SOCIAL_AUTH_GOOGLE_OAUTH2_SECRET= GOCSPCk6F8iMz7Zefz
FREE_MS_OAUTH=on
SOCIAL_AUTH_MICROSOFT_GRAPH_KEY= e0600780ae5241a8ef
SOCIAL_AUTH_MICROSOFT_GRAPH_SECRET= qKQ8QWdBRdzk
```
*Figure 6 - Example of FREE authentication configuration*

programmer also needs to write the JavaScript code that should be executed whenever a result is received from the server. This code receives the results in tabular form as formatted by the Control Experiment, converts them and updates the plots previously designed. This file should be placed in a specific directory in the FREE Server.

Each type of Apparatus, type of experiment and the specific physical apparatus requires the insertion of specific information on the database. This is done in the Django provided Administration interface (as illustrated in Figure 4**Error! Reference source not found.**). The FREE administrator should create one new entry in the Apparatus Type (corresponding to the characteristic of the specific physical Apparatus) and an entry in the protocol table. After this it is possible to create a new entry corresponding to the specific Apparatus.

The implemented API, the FREE Server organization and these installation steps allow the creation of new experiments without changes to the FREE Server Code (R12). Furthermore, the code necessary for the integration of a new experiment is limited in the number of lines of code and complexity.

*3) Services and libraries*

Besides the easy integration of new experiments, one of the main goals of developing a new RCL infrastructure was to allow its easy integration with existing external services such as authentication services or Learning Management Systems (requirement R10 and R11). This integration was eased by the selection of programming language and framework. Django already offers libraries that completely hide the implementation details and implement all the interoperability mechanism.

In FREE, the integration with existing authentication services (such as Microsoft, google, or other specific university authentication) is carried out using the OAuth [22] protocol, an industry standard that with a suitable client library only requires small configuration. In FREE the social-django [23] library is used to activate the OAuth protocol, this library takes advantage of the user management embedded in Django and with a simple configuration, the authentication for Microsoft or google services can be activated (Figure 6). The specific keys and secrets should be generated in the Cloud infrastructure administration portal. Specific university authentication providers can also be managed by this library but require the development of a python class.

Another integration that considered fundamental and already implemented was with external Leaning Management

*Figure 5 - FREE User Authentication a) login page, b) Users in database, c) external users.*

Systems (LMS) such as Moodle. To do this we selected the use of the industry standard LTI [20], by using the django-lti-provider [24] library it is now possible for Moodle (or any other LMS) to redirect users to FREE to execute any experiment. After being redirecting to FREE, users will be automatically logged in without requiring any further credentials. The LTI protocol handles and transfers user credentials and all complementary user information.

FREE also supports the streaming of video from the apparatus to the user browser (requirements R5). This is carried out using WebRTC [25] protocol: a camera attached to the Apparatus sends a video stream to the FREE server that, when asked, forwards it to the user browser. This decision allows the use of a pre-existing video streaming server (Janus [26] in current version of FREE) and relieves from the installation of video visualization software on the user computer.

## IV. VALIDATION

This section presents describes the functionalities and experiments that are now deploy and being used in real FREE installations.

### A. Ported experiments

Currently 6 different Protocols are ported to the FREE infrastructure. Two of these Protocols share the same Apparatus (as illustrated by Apparatus A in Figure 2). All the apparatuses were previously available through a different RCL infrastructure and provide experiments in physics:

- World Pendulum – Mechanics: Pendulum experiment devoted to study the local gravity dependence with latitude but used for other physics studies requiring such experiment;
- Inclined plane – Mechanics: This experiment is used as the typical experiment for the gravity determination with a small cart but equipped with a flap to increase the kinetic friction to model friction by numerical methods;
- Photovoltaic panel – Energy: A dimmed tricolored LED panel is used to produce electrical energy in a small solar panel. The user can do several studies according to the angle of incidence, irradiated power and wavelength, including the efficiency according to the impedance matching to the source.
- Langmuir Probe – Plasma Physics: This advanced experiment allows for the characterization of a plasma produced by RF. The user collects the I-V curve of a Langmuir probe from where we can extrapolate the electronic temperature and plasma density.
- Cavity – Plasma Physics-This advanced experiment is used for studying the electromagnetic propagation in a resonant cavity and observe how the existence of a plasma shifts the resonant frequency by altering the speed of light as soon as a RF plasma is generated.

Currently there are Pendulum apparatus and protocol connected to various FREE installations in Europe, Oceania, Africa, and Latin America.

The migration of Apparatus and protocols to the FREE infrastructure required a new implementation of the Experiment Control to use the available endpoints and data generated in the UI. This development was accomplished in a short amount of time and was done by a MSc Student with intermediate programming knowledge.

The effort to create the UI can be quantified by the number of lines of code necessary to define the whole UI. For an experiment whose configuration has 2 numeric values, and the results are presented in one table and 3 plot the programming effort is as follows:

- HTML input form - 20 lines for the input of 2 values
- HTML results presentation - 25 lines for one table and 3 plots
- JavaScript code – 180 lines for the update of the table and plots

The effort to create the UI (HTM and JavaScript) for a new experiment (after the implementation of the corresponding Experiment Control) and the structure of the Execution configuration defined is, on average, lower that 5 hours of continuous work.

### B. Remote Services

The FREE allows the user authentication using external services such as Google, Microsoft or even University services, besides locally created users. Currently the running installations integrate these external services differently.

Figure 5.a) shows the login page, where the user can either input the FREE username and password (for administrators, for instance) or select and external authentication service.

In this FREE installation three external services are configured that automatically appear on this page. After selecting one of these providers the user is redirected to the corresponding login page, and after successfully authentication, the user can access FREE as any other user. Figure 5.b) and c) show the user table where the first two users where authenticated using the Google and the university services. The wp-guest and wp-admin are local users that have one local password.

The third entry from Figure 5.b) was transparently created when the user accessed FREE from the Moodle installation.

Before professors can provide access to FREE experiments in their courses, LMS administrator needs to register the FREE installation (**Error! Reference source not found.**a). The administrator only needs to insert the network address of the FREE installation (Tool URL), the Consumer key and Shared Secret. This information is responsibility of the FREE administrator to provide. After this configuration, professors can create activities in their courses that will redirect students into FREE. During the course, the student clicks on the provided activity (Figure 7.b) and is redirected into FREE, as occurred with the student@university.tv in Figure 5.b). Although only evaluated with Moodle, the use of LTI allows the integration of FREE into most LMS following similar steps.

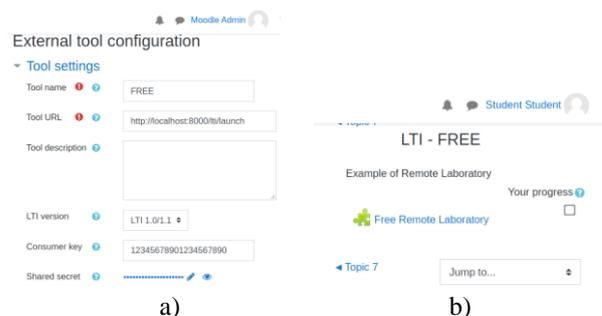

*Figure 7 - LTI integration into FREE a) registration of FREE into Moodle b) Student access in Moodle*

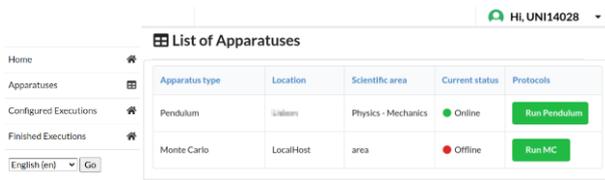

*Figure 8 - FREE interactions a) main menu, b) list of available apparatuses*

If using these authentication mechanisms, any FREE installation becomes public and with universal access. To guarantee a fair use of the resources, experiment usage policies and mechanisms to delete old and inactive users were implemented into FREE. This way users will not be able to abuse the systems, the number of database resources will be limited in order not to degrade other users´ experience.

*C. Functionalities*

After a user enters FREE a home page with a simple menu is presented (**Error! Reference source not found.**.a) allows the access to installed Apparatus and Protocols, Configure execution and results of past executions. FREE is multigoal with translations to English, Portuguese, and Spanish.

The page (Figure 8.b) with the list of the installed apparatus, presents basic information about each apparatus, the available protocols and it state. From this page users can create a new experiment pressing the Run button.

The experiment configuration page presents three tabs, one with the description of the experiment, another with the configuration form and a final one with results of the execution. A small window with the video stream from the experiment apparatus allows the user to see in real time the execution of the experiment.

One example of a Experiment tab is illustrated in Figure 9. This corresponds to the configuration of the Pendulum Experiment, where the used must define the initial pendulum displacement and the number of samples to be measures. The three buttons allow the user to save the configuration reset it or start the execution.

After being save, an experiment configuration can later be accessed and modified. Only when the Experiment configuration is submitted it may be executed by the apparatus.

During the execution of the experiment the user can see the video in real time, but also see the plot and table being updated (Figure 10).

The table with all the relevant experiment measurements can be downloaded in various formats (excel, or csv) for latter processing. The plot are dynamic allowing the user to zoom in specific or highlight parts of the data.

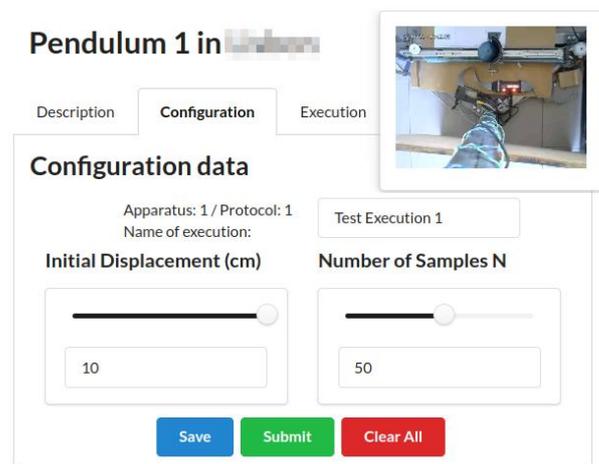

*Figure 9 - Experiment configuration window*

In the development of this page, the experiment programmer only had to define the two input fields and sliders. All the logic related to the buttons, save of the configurations, and start of the experiment is generic and handled by the FREE server.

Although the FREE user interface is minimal in the number of pages (home pages, list of apparatuses, list of user saved experiments, list of user experiment executions results, and Execution page) the fundamental functionalities available on currently existing RCL systems are already implemented in this first version, fulfilling the basic functional requirements R1 to R6.

V. CONCLUSION

FREE is a new RCL system that, by using modern programming languages, techniques and software architectures, allows the simple development of remote controlled laboratories. FREE provides a portal to experiments that can be scattered in various institutions of networks and allows their integration and access through external authentication services and existing LMSs.

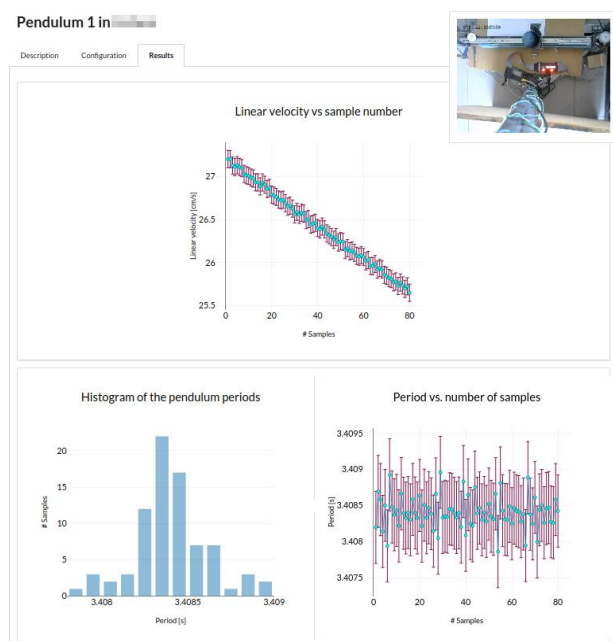

*Figure 10 - Experiment execution results*

The architecture, the explicit separation of components and the way they interact guarantees that the development or adaptation of new experiments into FREE is done is a fast and efficient way. The programming skills necessary to do these developments is also low. Only basic proficiency in python and web programming is enough.

FREE allows the development and integration of remote experiments developed in a programming language different from python, because the remote interactions are well defined and based on REST web services. This explicit separation also allows the development of virtual experiments whose access can be mediated by FREE. In these case the Experiment Controller of the virtual experiment would simulate the physical process and be integrated with FREE as any remote experiment.

FREE is currently being developed and extended by a heterogenous small programming team ranging from Electrical and physics engineering students to one already graduated Informatics Engineers. In the future this interdisciplinarity can be further leveraged in the development of new functionalities (such as Apparatus reservation or execution priorities), or new types of interactions, that require the knowledge and expertise in algorithms, computer engineering and Informatics. The development of new experiments can be done in the future in the context of MScs in specific study fields (physics, EEC, biology, chemistry, or others).

The integration with LMSs can also be further extended. Now besides redirecting students to available experiments no other functionality is implemented. At the moment there is a Computer Engineering student implementing mechanism to allow the definition of questions (multiple choice, or numeric answers) whose responses depend on the execution and results of a real experiments. Here, students will execute a real remote experiment, process the data, and answer questions that will be automatically assessed.

Another future possibility lies in the creation of a federation of RCL, where one installation of FREE in a certain institution, will provide seamless and transparent access to experiments in remote FREE installations. This approach will foster communication and collaboration in the development and deployment of experiments, but also increase the availability of resources (number and diversity of experiments) to users.

From the implementation efforts and attained results, we think that FREE successfully handles the limitations identified in current systems, but also provides a starting platform for the evolution, dissemination and increase of user of remote controlled laboratories.